\documentclass[journal]{IEEEtran}

\usepackage{epsfig,amsmath,amssymb,epsf,amsthm,scalefnt,multirow,subfig}
\usepackage{xcolor}
\usepackage{float}
\usepackage{cite}
\usepackage{psfrag}
\usepackage{booktabs}

\newtheorem*{lemma*}{Lemma}

\def\b0{{\pmb{0}}} 

\begin{document}
\bstctlcite{IEEEexample:BSTcontrol}

\title{Millimeter Wave Vehicular Communication to Support Massive Automotive Sensing}

\author{Junil Choi, Vutha Va, Nuria Gonz\'{a}lez-Prelcic, Robert Daniels, Chandra R. Bhat, and Robert W. Heath Jr.\\
\thanks{J. Choi, V. Va, R. Daniels, and R. Heath are with Wireless Networking and Communications Group, The University of Texas at Austin, Austin, TX 78712, USA (email: \{junil.choi,vutha.va,robert.daniels,rheath\}@utexas.edu).}
\thanks{N. Gonzalez-Prelcic is with Universidade de Vigo, Vigo, Spain (email: nuria@gts.uvigo.es).}
\thanks{C. Bhatt is with Center for Transportation Research, The University of Texas at Austin, Austin, TX 78712, USA (email: bhat@mail.utexas.edu).}
\thanks{This research was partially supported by the U.S. Department of Transportation through the Data-Supported Transportation Operations and Planning (D-STOP) Tier 1 University Transportation Center and by the Texas Department of Transportation under Project 0-6877 entitled ``Communications and Radar-Supported Transportation Operations and Planning (CAR-STOP)''.}}

\maketitle

\begin{abstract}
As driving becomes more automated, vehicles are being equipped with more sensors generating even higher data rates. Radars (RAdio Detection and Ranging) are used for object detection, visual cameras as virtual mirrors, and LIDARs (LIght Detection and Ranging) for generating high resolution depth associated range maps, all to enhance the safety and efficiency of driving. Connected vehicles can use wireless communication to exchange sensor data, allowing them to enlarge their sensing range and improve automated driving functions. Unfortunately, conventional technologies, such as dedicated short-range communication (DSRC) and 4G cellular communication, do not support the gigabit-per-second data rates that would be required for raw sensor data exchange between vehicles. This paper makes the case that millimeter wave (mmWave) communication is the only viable approach for high bandwidth connected vehicles. The motivations and challenges associated with using mmWave for vehicle-to-vehicle and vehicle-to-infrastructure applications are highlighted. A high-level solution to one key challenge --- the overhead of mmWave beam training --- is proposed. The critical feature of this solution is to leverage information derived from the sensors or DSRC as side information for the mmWave communication link configuration. Examples and simulation results show that the beam alignment overhead can be reduced by using position information obtained from DSRC. 
\end{abstract}

\section{Introduction}\label{sec1}
\subsection{Motivation}

The number of sensors on vehicles, and the rate of data that they generate, is increasing. The average number of sensors on a vehicle today is around 100, but that number is expected to double by 2020 as vehicles become smarter and automated \cite{Lu:2014}. At present, automotive radars and visual cameras are the most common safety sensors found in vehicles \cite{Meinel:2013,Fleming:2012}. By sensing the existence, position, and velocity of other vehicles or objects, automotive radars make it possible to realize adaptive cruise control, blind spot detection, lane change assistance, parking assistance and more. Cameras make better driving environments, e.g., eliminate blind spots and work as virtual mirror, and provide better night vision with infrared sensors. Further, recent developments on autonomous vehicles heavily rely on LIDARs, which use laser technology to generate high resolution depth associated range maps. The amount of data generated by LIDARs is similar to conventional automotive cameras, which will further increase the amount of data generated by a vehicle. Active safety algorithms therefore need to work with more sources of data and higher data volumes.



A major challenge associated with many sensor technologies is that they have a limited sensing range.  For example, radar, cameras, and LIDARs provide information only about objects within the line-of-sight of the sensor, which limits the automation capability of vehicles for better safety.  An alternative is to employ wireless communications to enable cars to exchange information to create what is known as a connected vehicle. Various safety-related applications with improved automation capability are enabled by vehicle connectivity (even with very limited information provided by current connected vehicle technologies), including forward collision warning, do not pass warning, blind intersection warning, and red light violation warning, which may reduce more than 80\% of all annual car crashes \cite{Harding:2014}. Vehicle connectivity has two potential benefits. First, if a suitable carrier frequency is chosen, then cars can communicate in non-line-of-sight conditions, for example around corners. Second, if a high bandwidth communication link is available, cars can exchange higher rate raw sensor data (with minimal preprocessing), which we call \textit{fully connected vehicles} in this paper. Fully connected vehicles can implement powerful active safety applications, e.g., ``See Through'' and ``Bird’s Eye View'' identified in the 5G-PPP white paper on the automotive vertical sector.\footnote{The white paper can be found in https://5g-ppp.eu/wp-content/uploads/2014/02/5G-PPP-White-Paper-on-Automotive-Vertical-Sectors.pdf.} The shared raw sensor data can be independently processed by each vehicle (as today’s prototype autonomous vehicles do). By sharing raw sensor data, it becomes possible to implement adaptive platooning or even cloud-driven, fully automated driving (if processing can be done in an edge cloud to have acceptable latency). This has the potential to further improve transportation efficiency and reliability.

\subsection{Related work and proposed approach}
The state-of-the-art protocol for connecting vehicles is called dedicated short-range communication (DSRC) \cite{Kenney:2011}. Using DSRC, it is possible to implement preliminary vehicle-to-vehicle (V2V), vehicle-to-infrastructure (V2I), or even vehicle-to-everything (V2X) communication systems.  The National Highway Traffic Safety Administration (NHTSA) is likely to mandate that all new vehicles include DSRC capability by 2017 \cite{Harding:2014}. While DSRC permits vehicles to exchange messages (including basic sensor information) with a range up to 1000 meters (ideally), the maximum data rate in practice is only 2-6 Mbps. Fourth generation (4G) cellular systems using device-to-device (D2D) mode could be used for V2X communication systems, which is currently being discussed in the 3GPP standard \cite{Seo:2016}. The maximum data rate of 4G systems, however, is still limited to 100 Mbps for high mobility, though much lower rates are typical. Therefore, current technologies are not sufficient to handle the terabyte-per-hour data rates that can be generated in next generation vehicles. 

In this paper, we motivate the use of millimeter wave (mmWave) spectrum, on top of DSRC or 4G V2X, for fully connected vehicles. MmWave is already available to consumers in the form of IEEE 802.11ad \cite{IEEE11ad:2012} and is likely to be a part of the fifth generation (5G) cellular systems. The use of mmWave provides access to high bandwidth communication channels, leading to the potential for gigabit-per-second data rates and realizing raw sensor data exchange among vehicles and infrastructure. In fact, mmWave is not new to the automotive world. Current automotive radars already use mmWave spectrum \cite{Meinel:2013}. For vehicular communications, mmWave has been tested more than a decade ago \cite{Kato:2001}. There is even a standard for vehicular communications at mmWave from the International Organization for Standardization (ISO) \cite{ISO:2013}, although it is not a complete system; it only provides several parameter values of the physical layer. Therefore the foundations for mmWave communication in V2X are already available, and we believe it is now the time to revisit the use of mmWave for vehicular communications considering much interest in vehicular automation.

\begin{table*}[t] \centering
\caption{Purposes, drawbacks, and data rates of automotive radar, camera, and LIDAR. The sensor data rates are obtained from the specs of commercial products and conversations with industrial partners.}\label{sensing_comp}
\begin{tabular}{|c||c|c|c|}
  \hline
             & Purpose & Drawback & Data rate \\
    \hline
	Radar & Target detection, velocity estimation & Hard to distinguish targets & Less than 1 Mbps \\
	\hline
	\multirow{2}{*}{Camera} & \multirow{2}{*}{Virtual mirrors for drivers} & \multirow{2}{*}{Need computer vision techniques} & 100-700 Mbps for raw images,\\
	& & & 10-90 Mbps for compressed images\\
	\hline
	\multirow{2}{*}{LIDAR} & Target detection and recognition, & \multirow{2}{*}{High cost} & \multirow{2}{*}{10-100 Mbps} \\
	& velocity estimation & &\\
	\hline
\end{tabular}
\end{table*}

We begin this paper by reviewing different automotive sensing techniques that are frequently used for active safety functions. The discussion shows why high data rates are required to exchange raw sensor data between vehicles. We explain that these high data rates could be provided through the use of 5G mmWave cellular, a modification of IEEE 802.11ad, or the development of a dedicated mmWave V2X networking protocol. We review the three main challenges to implement mmWave V2X communication systems, i.e., the lack of mmWave vehicular channel model, the penetration rate of mmWave V2X-capable vehicles, and mmWave beam alignment. Then, we focus on the beam alignment problem, which will be difficult in vehicular environments because of high mobility. We propose a new way to reduce the overhead for beam alignment by exploiting DSRC or sensor information as side information. We provide specific examples where position information obtained from DSRC is used to reduce the beam alignment and tracking overhead in mmWave V2X scenarios. 

\section{Sensing for Next Generation Vehicles}\label{sec2}
\begin{figure}[t]
  \centering
  \includegraphics[width=1\columnwidth]{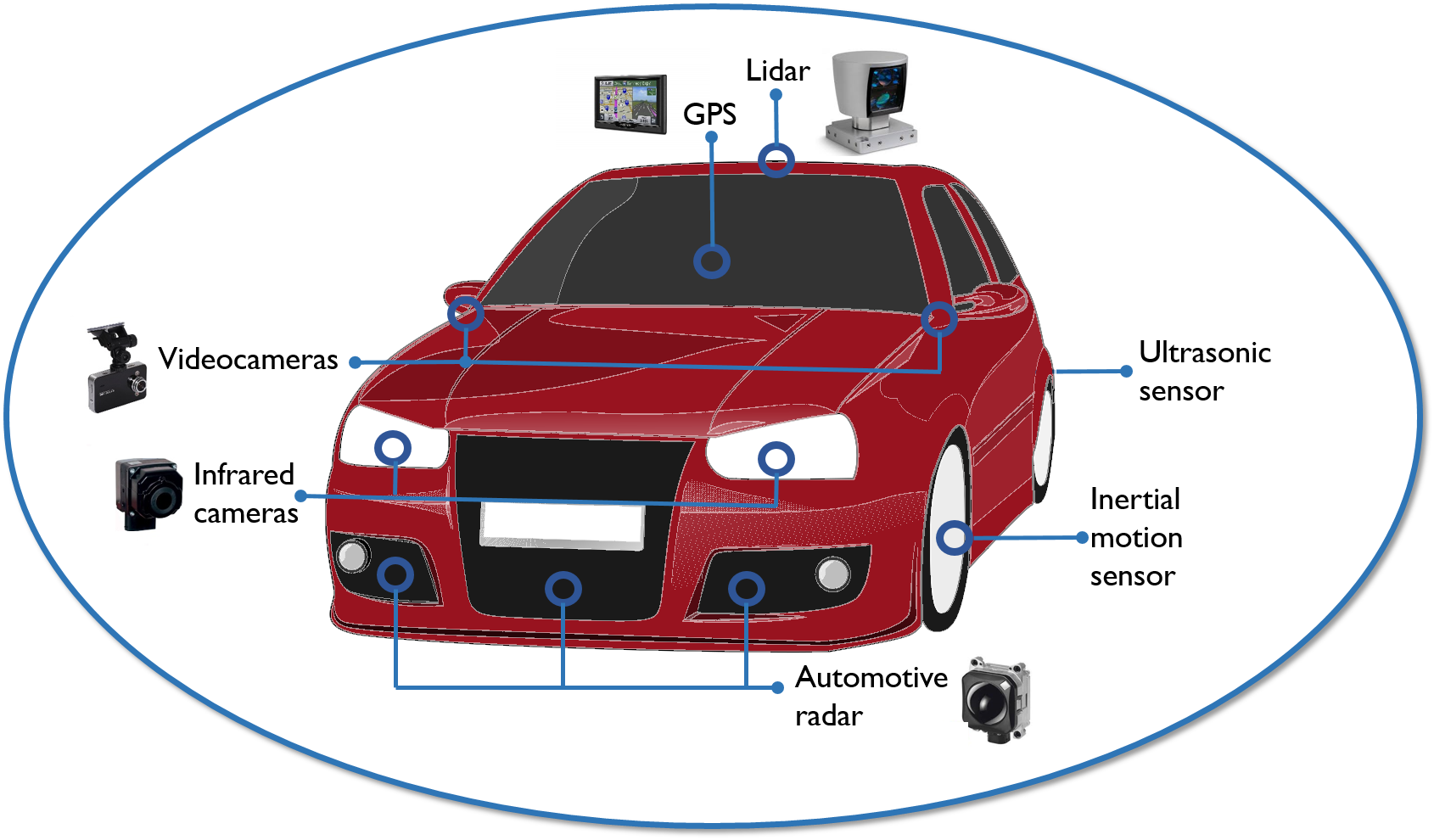}\\
  \caption{A car with many safety-related automotive sensors.}\label{car_sensor}
\end{figure}

In this section, we discuss the purpose and characteristics, drawbacks, and the amount of data generated from three major automotive sensors, i.e., radars, cameras, and LIDARs, for current and next generation vehicles, which are illustrated in Fig. \ref{car_sensor}. The discussion is summarized in Table \ref{sensing_comp}.

Automotive radars are already operating in mmWave spectrum, most recently between 76 and 81 GHz frequencies \cite{Meinel:2013}. Long range radars are used for adaptive cruise control. Medium range radars support cross traffic alerts, lane change assistance, and blind spot detection. Short range radars help assist in parking and pre-crash applications. Radars engage in active sensing by sending a special waveform, typically a frequency modulated continuous waveform, to derive information about the existence, location, and velocity of neighboring vehicles from reflections. The waveforms and antenna structures vary by manufacturer unlike communication systems that are typically standardized. The European Telecommunications Standards Institute (ETSI) has defined some waveform-related parameters, e.g., spectrum, peak and average allowable power, and out-of-band emission, but the overall waveforms and signal processing are not specified. Radars are not suitable to recognize the types of target objects (i.e., whether the detected target is a truck, sedan, or motorcycle) without modifying their standard signal processing. The data is heavily post processed to reject clutter and produce point-map information only for relevant targets. Data rates generated by radar vary from kbps (for point-maps) to hundreds of Mbps (for raw sampled data).

Automotive cameras use visible or infrared spectrum. They provide the driver with an additional view for safety, e.g., front cameras detect speed limit signs and improve night vision (with infrared sensors), side-rearview cameras check blind spots and lane departure, rearview cameras prevent backover crashes, and interior cameras prevent the driver from dozing off. New computer vision algorithms are required for automotive cameras to work as (intelligent) safety applications. The amount of data generated from automotive cameras is large. Typically, the data rates vary from 100 Mbps for low-resolution to 700 Mbps for high-resolution raw images at 15 Hz frame frequency, with up to a 10x reduction due to compression. Automotive cameras form one of the highest data rate sources of sensor data on the car.

LIDARs use narrow laser beams to generate high resolution three-dimensional (3D) digital images, where pixels are also associated with depth, and with a 360 degree field of view by a rotating laser. LIDAR operates by scanning a laser over an area and measuring the time delay from the backscattered signal.  With suitable post processing, LIDAR can image with very high resolution and can be used for 3D maps as well as for car, bicycle, and pedestrian detection.  This has made them a key feature of Delphi's and Google's autonomous vehicles. While they are costly at present, e.g., \$8,000-\$80,000 per device, the costs are expected to drop in the future. LIDAR data rates are comparable to those of automotive cameras, i.e., 10-100 Mbps depending on specifications, e.g., the numbers of laser beams and images per second.

\section{MmWave V2X Communications}
In this section, we first discuss general mmWave communication systems. Then, we sketch out potential V2V and V2I architectures implemented using mmWave to support fully connected vehicles that share their raw sensor data. We also explore the possibility of realizing these architectures using 5G, a modification of IEEE 802.11ad, or a dedicated new standard.
\begin{figure}[t]
  \centering
  \includegraphics[width=1\columnwidth]{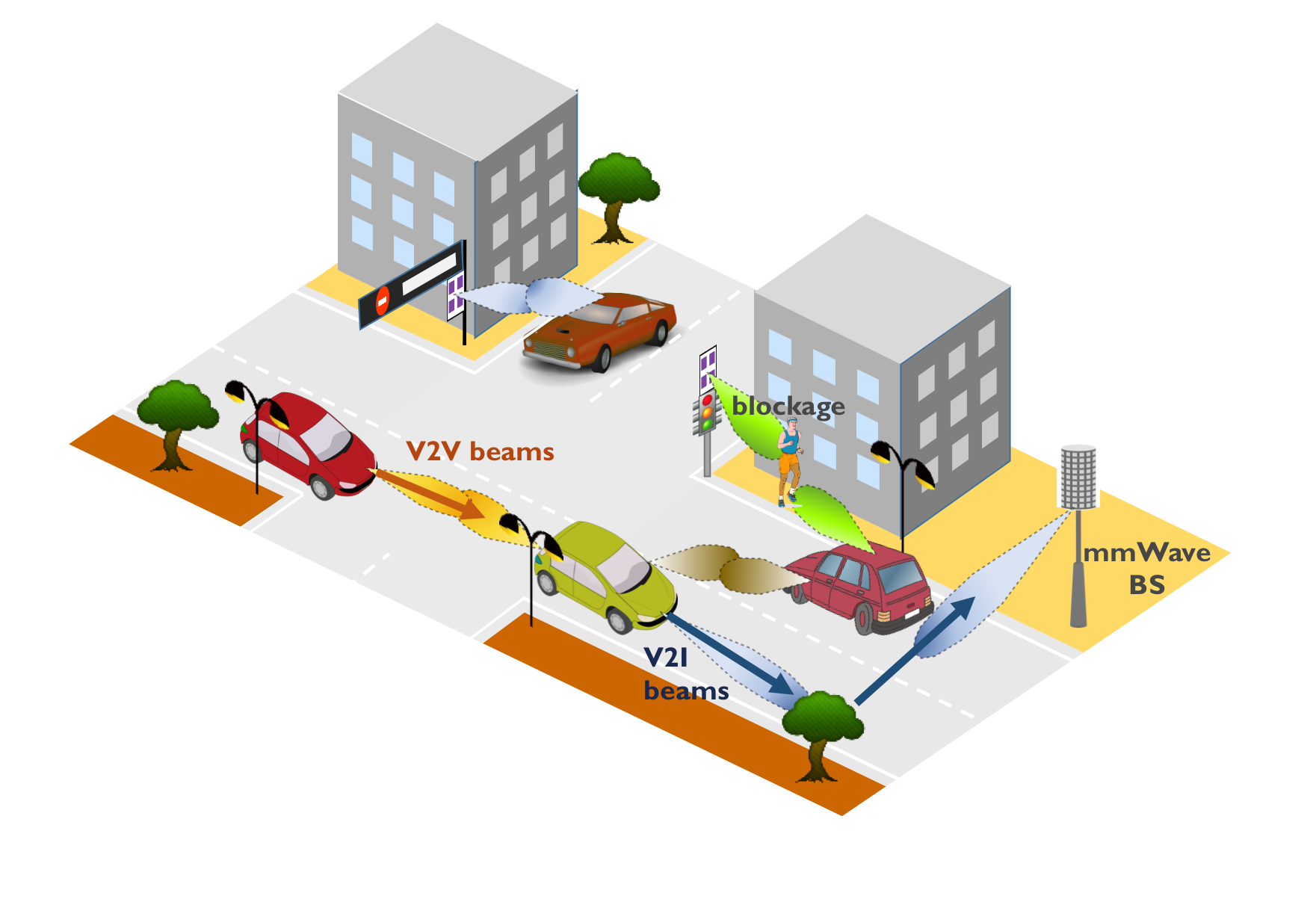}\\
  \caption{The conceptual figure of V2X communications using mmWave. Multiple mmWave transceivers are deployed on a vehicle to simultaneously establish V2V and V2I communication links.}\label{V2X}
\end{figure}

MmWave refers to the spectrum between 30 and 300 GHz. Using mmWave carrier frequencies for communication, it is possible to exploit larger spectral channels. Combined with higher order modulations and multiple-input multiple-output (MIMO) techniques that enhance spectral efficiency, mmWave can achieve higher data rates than those provided by current wireless communication systems operating at below 6 GHz carrier frequencies \cite{Roh:2014}. For example, IEEE 802.11ad uses 2.16 GHz of bandwidth in the unlicensed 60 GHz band and supports data rates up to 7 Gbps \cite{IEEE11ad:2012}. Recently, the FCC has proposed to authorize the operation in the 28, 37, and 39 GHz of the licensed band and made 64-71 GHz available for unlicensed spectra for mobile use, which would facilitate the use of mmWave for various wireless communication scenarios including V2X. 

In mmWave communications, it is essential to have a large number of antennas at the transmitter and the receiver to form sharp transmit and receive beams and establish good communication links. Due to small wavelengths of mmWave frequencies, it is possible to deploy a large number of antennas in a small form factor. Because of the large number of antennas, however, it may not be cost efficient to have high-quality signal processing components for all antennas. Therefore, analog beamforming (with one RF chain) and hybrid beamforming (with a few RF chains where the number of RF chains is far less then that of antennas) have drawn a significant interest for mmWave communication systems.  There has been much work demonstrating that the performance of hybrid beamforming is similar to that of full digital beamforming. We refer to \cite{Heath:2015} for details about general mmWave communication technologies.

Fig. \ref{V2X} illustrates mmWave V2X communications as a conceptual level. For mmWave V2X communications, there should be multiple mmWave transceivers to overcome the blockage of mmWave signals by nearby vehicles or even pedestrians. For example, a vehicle may have mmWave transceivers on front and rear bumpers and sides for V2V, and on its rooftop for V2I communications, because infrastructure will be placed in high positions to ensure good link conditions. The blockage effect in the mmWave V2V scenario might be mitigated by using the gap under vehicles as waveguard, which allows  a vehicle to communicate with vehicles other than adjacent ones. Although blockage is usually considered a defect, it can be beneficial for V2V communications because the effect (combined with narrow mmWave beams) can reduce inter-vehicle interference and enable better spatial packing, as demonstrated by using directional antennas in lower frequencies in \cite{Guha:2012}. In fact, blockage may not be a serious problem in mmWave V2X because only the vehicles in proximity would share their raw sensor data while the basic safety information such as position and velocity can be broadcast using DSRC or 4G V2X.

Multiple mmWave transceivers on a vehicle will likely be connected to the central controller of a vehicle via controller area network (CAN) bus (or more advanced in-vehicle network protocols that can support high data rates) to exploit high data rate information from neighboring vehicles. To avoid significant cable loss at mmWave frequencies, both RF and some baseband processing components should be integrated into a single mmWave transceiver. For example, the medium access control (MAC) protocol can be divided into two layers, i.e., distributed lower MACs at each transceiver connected to a centralized upper MAC by lower-rate CAN bus.

5G cellular systems may use mmWave spectrum to provide gigabit-per-second data rates. It is possible that mmWave V2X communications are supported through 5G communication systems, i.e., 5G base stations serve as infrastructure for V2I communications, and 5G D2D mode supports V2V communications. To help set up mmWave links, a low-frequency control plane (using DSRC or 4G cellular) can be exploited. In this way, it is possible to leverage existing cellular infrastructure and the ability of mass production of consumer cellular devices to manufacture mmWave V2X transceivers. Using 5G for V2X communications, however, needs a clever strategy to share the spectrum with many other cellular devices and must support the high mobility. 

As part of DSRC, i.e., IEEE 802.11p evolved from WiFi standards, it seems reasonable to exploit IEEE 802.11ad for mmWave V2X communications. IEEE 802.11ad was, however, originally designed for indoor use and the standard needs some modifications to support high mobility. For example, hierarchical beam sweeping in IEEE 802.11ad may work for indoor environments, but it is questionable whether the technique would work for high mobility scenarios. The MAC protocol also should be revisited to support intermittent connectivity in vehicular environments. DSRC may give insight on adapting IEEE 802.11ad to vehicular environments.

Instead of relying on 5G systems or existing WiFi standards, it is also possible to have a new standard (or a revised version of the ISO standard \cite{ISO:2013}) with dedicated spectrum for mmWave V2X communications similar to DSRC. Having a separate mmWave V2X standard will enable flexibility to optimize vehicular communication systems while requiring additional efforts on manufacturing mmWave V2X equipment. 

\section{Challenges for mmWave in V2X}

There are many challenges to develop mmWave V2X communication systems. In this section, we focus on three main problems: (i) lack of accurate mmWave vehicular channel models, (ii) required penetration rate of mmWave V2X-capable vehicles, and (iii) lack of simple and fast mmWave beam alingment algorithms for vehicular communications. 

\subsection{MmWave vehicular channel modeling}
Wave propagation between vehicles has been studied at frequencies below 6 GHz, and different channel models have been proposed \cite{Mecklenbrauker:2011}. Vehicular channel models used at these frequencies include geometry and non-geometry based stochastic models, ray tracing approaches, and graph based models; all these approaches have to be modified and extended to mmWave bands, inferring the new parameters from the measurement data. There are also some propagation and angle-of-arrival measurements at 60 GHz \cite{Kato:2001,Dor:2011}. Further measurements are needed at mmWave bands to characterize the impulse response between antenna arrays in vehicular scenarios. The effects of the location of antennas and the blockage caused by nearby vehicles, buildings, and pedestrians must be measured and incorporated into mmWave V2X communication channels.

\subsection{Penetration rate of mmWave V2X-capable vehicles}
In the early stage of deploying mmWave V2X communication systems, the insufficient penetration rate of V2X-capable vehicles would be an issue to exploit the full benefit of V2X (especially V2V) communications. High penetration rates may cause excessive interference in highly-loaded traffic conditions, which could be mitigated with narrow mmWave beams and advanced MAC protocols. Because automotive radars already exploit mmWave spectrum, it may be possible to implement joint mmWave radar and communication systems and increase the penetration rate of mmWave V2X-capable vehicle in the early deployment stage. Joint systems will also save space in a vehicle and reduce cost and power consumption. A preliminary study on designing a joint mmWave radar and communication framework for vehicular environment based on the IEEE 802.11ad system (operating in the 60 GHz unlicensed spectrum) was conducted in \cite{Kumari:2015}.  It was shown in \cite{Kumari:2015} that a range estimation accuracy within 0.1m and a velocity estimation accuracy within 0.1m/s could be achieved by using the IEEE 802.11ad preamble as a radar signal and standard WiFi receiver algorithms. It may also be possible to add communication functions on existing mmWave automotive radars operating in 76-81 GHz frequencies. 

\subsection{MmWave beam alignment overhead}
Due to common use of sharp transmit and receive beams in mmWave systems, the beam alignment between the transmitter and the receiver is critical in mmWave communications. Once beams are properly aligned, standard communication protocols, i.e., effective channel estimation and data transmission, can be performed using sufficient link margin. The brute force approach to perform beam alignment is testing all possible transmit and receive beams sequentially, which has high overhead. There has been much work on reducing the beam alignment overhead in cellular and WiFi systems. MmWave V2X communications, however, may require much advanced beam alignment techniques considering the high mobility of vehicles. MmWave V2I and V2V communications may require different approaches for beam alignment. Because the infrastructure is stationary, the fast initial beam alignment would be important in mmWave V2I communications. In mmWave V2V communications, beam tracking after the initial beam alignment would be more important because neighboring vehicles are likely to be moving in similar speeds. In both mmWave V2I and V2V cases, beam realignment caused by blockage should be properly handled.


\section{V2X MmWave Beam Alignment using DSRC or Automotive Sensors}
In \cite{Guha:2012}, using sharp transmit and receive beams (or directional antennas) in low frequencies has been proposed to improve packet deliver ratios or latency; however, the beam alignment problem has been neglected by assuming a quasi-stationary network. We investigate possible ways to mitigate the beam alignment overhead in this section. The key idea is to use automotive sensors or DSRC to obtain relative position information of neighboring vehicles. Due to the high directivity of mmWave channels, this relative position information and trajectory estimation performed by vehicles and infrastructure can be exploited as side information to mitigate mmWave beam alignment and tracking overhead. The notion of using out-of-band sensing techniques has not been received much interest in wireless communications industry so far because of lack of necessity (training overhead may not be a problem) or lack of available sensors. In mmWave V2X, however, exploiting out-of-band sensors is sensible considering (i) the large overhead of establishing and maintaining communication links and (ii) the availability of many sensors.

Vehicles are able to obtain other vehicles' relative position information using automotive sensors. To obtain accurate position information, a vehicle may need to combine the information from multiple sensors. For example, a vehicle can detect the existence, relative positions (e.g., distance and angle from the vehicle), and velocity of some objects using radar. It may be difficult for radars to tell whether the detected objects are other vehicles, obstacles, or infrastructure. On the contrary, automotive cameras can tell possible vehicles and infrastructure for communications using computer vision techniques; however, the distance information obtained from cameras is typically less accurate. LIDARs that have characteristics of both radars and cameras may ease the problem, albeit with higher cost. Vehicles can also predict other vehicles' trajectory, which can be used for beam tracking, using their sensors information. 

It is also possible to use DSRC signals from neighboring vehicles to reduce the beam alignment overhead because DSRC signals such as basic safety messages (BSMs) contain useful information, e.g., absolute position, velocity, size, and heading of neighboring vehicles, for beam alignment. A vehicle can deduce relative position information of neighboring vehicles by using its absolute position (from its own GPS) and the absolute position information from neighboring vehicles. Using this information, vehicles or infrastructure would know the relative position of vehicles with which they want to establish mmWave communication links. Infrastructure can also broadcast its absolute position information to incoming vehicles using DSRC signals for mmWave V2I beam alignment.

Even with relative position information from automotive sensors or DSRC, vehicles still need to perform beam alignment because (i) the absolute position information of a vehicle may be inaccurate due to GPS estimation noise or insufficient satellite visibility, and (ii) a vehicle may not know the exact position of mmWave transceivers of other vehicles. When performing beam alignment, each vehicle can adaptively decide proper candidate beams for beam alignment by using other vehicles' relative position and size information obtained from its sensors or DSRC signals from other vehicles. Once mmWave communication links are established, the necessary information for beam tracking can be exchanged with high data rates of mmWave links. 

\section{Evaluation on mmWave V2X Beam Alignment}

In this section, we show example scenarios where the proposed idea of using DSRC or automotive sensors is beneficial to reduce the mmWave beam alignment overhead. Then, we evaluate the proposed idea with numerical results considering practical vehicular environments.
\begin{figure}[t]
  \centering
  \includegraphics[width=1\columnwidth]{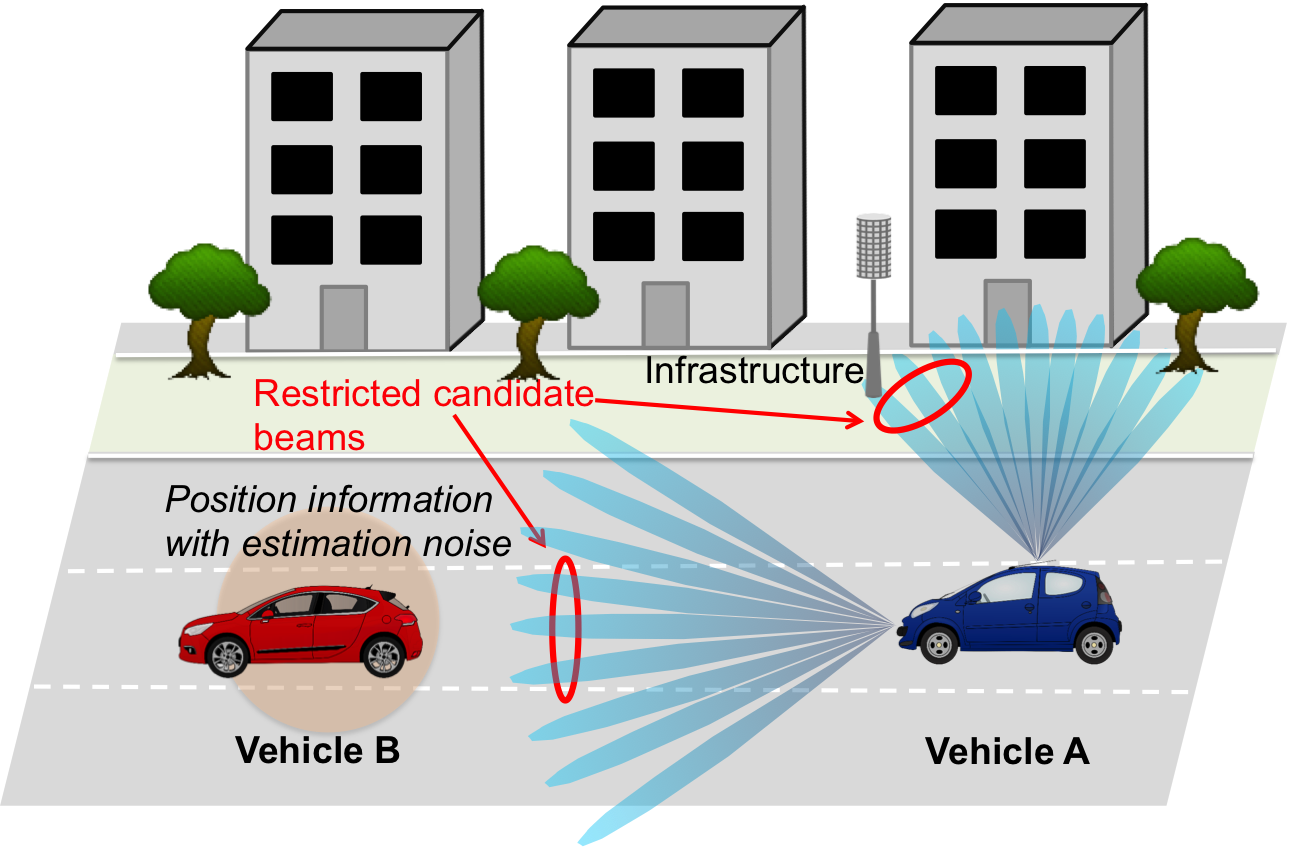}\\
  \caption{Illustrative examples of the proposed idea of using position information from DSRC or automotive sensors to reduce the mmWave beam alignment overhead.}\label{example}
\end{figure}

\subsection{Illustrative examples}
\begin{figure*}
\centering
\subfloat[A snapshot of ray-tracing simulations.]{
\includegraphics[width=1\columnwidth,height=200pt]{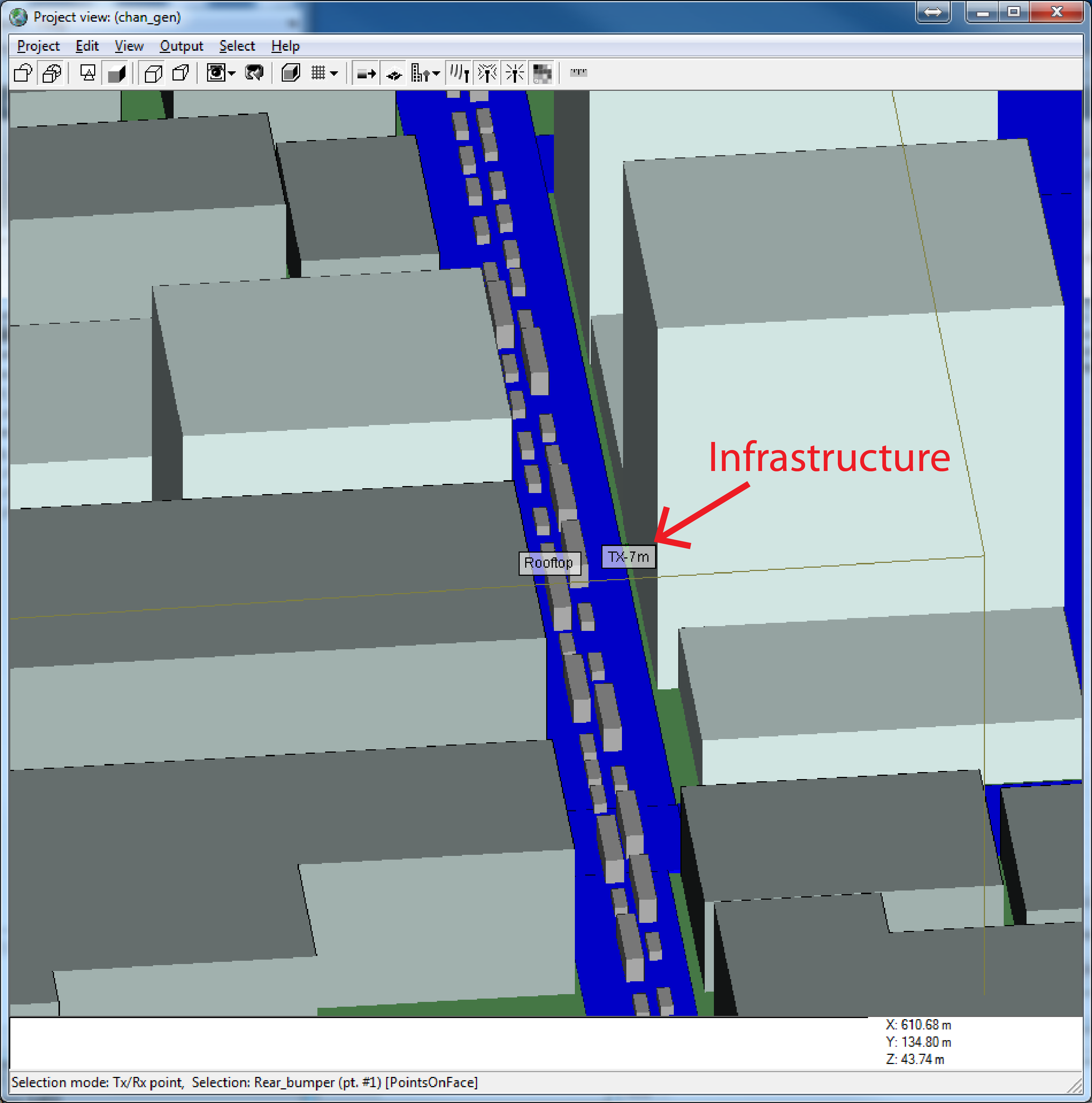}
\label{raytracing}
}
\subfloat[3D beam patterns of the $8\times 8$ UPA.]{
\includegraphics[width=1\columnwidth,height=200pt]{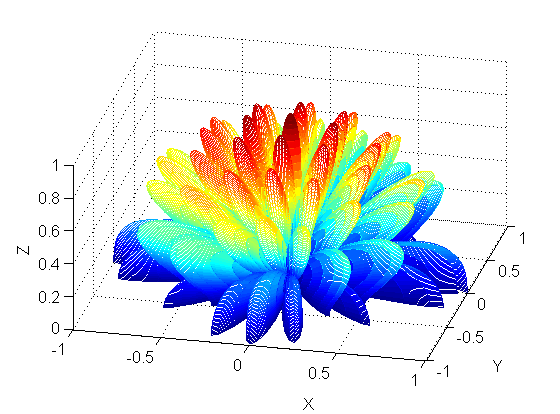}
\label{beampattern}
}
\caption{Considered mmWave V2I scenario and the 3D beam patterns of the training beams constructed by the Kronecker-product of DFT vectors.}
\label{simul_setup}
\end{figure*}

Consider a V2V communication scenario where a vehicle~A wants to establish a communication link with another vehicle~B in front as shown in Fig. \ref{example}. Assume vehicle A knows the existence of vehicle B using automotive sensors but does not know the exact relative position of vehicle B. To establish a mmWave communication link by beam alignment in this case, vehicle A may need to test many beams to select the best beam for communications. Considering the narrow beams in mmWave communications, the beam alignment overhead will be quite large. It is possible to exploit advanced beam alignment techniques, e.g., hierarchical beam sweeping approach in IEEE 802.11ad \cite{IEEE11ad:2012}, to reduce the beam alignment overhead. We believe though that the fast moving nature of vehicles will require novel beam alignment techniques for V2X. 

If vehicle A knows the relative position of B in advance by using DSRC signals or automotive sensors, it is possible for A to restrict the candidate beams for the beam alignment. The restricted beam search space will effectively reduce the beam alignment overhead without any performance loss. Advanced beam alignment techniques can be exploited on top of the restricted beam search space to further reduce the beam alignment overhead. 

In the V2I case, the benefit of using position side information for mmWave beam alignment will be even greater. Usually infrastructure will be located in the side of roads, which leads to a wider angular region for searching the best beam than in the V2V case. Because of the different heights of vehicles and infrastructure, 3D beamforming that controls beam patterns in both azimuth and elevation will be necessary for mmWave V2I communications, which also increases the beam search space. Since infrastructure is stationary, it may have prior knowledge of its neighboring environments, e.g., preferred beams for a specific region, which can be further exploited. Using DSRC signals from infrastructure or accurate real-time 3D map data, vehicles entering into the serving area of infrastructure may know the position of the infrastructure in advance and can effectively restrict the beam search space. 

\subsection{Numerical evaluation}

We evaluate the proposed idea of exploiting DSRC to reduce mmWave beam alignment overhead by using a ray-tracing simulator to mimic realistic road environments. We consider a V2I scenario where infrastructure and a vehicle (which is called as the communicating vehicle (CV) throughout the section) are trying to establish a mmWave communication link using 60 GHz carrier frequency through beam alignment. A snapshot of ray-tracing simulations is depicted in Fig. \ref{raytracing}. The distance between two consecutive vehicles follows the Erlang distribution, and two types of vehicles are considered: (i) a vehicle with the size of $5\times 1.8\times 1.5$ (m$^3$) and (ii) a truck with the size of $12\times 2.5\times 3.8$ (m$^3$). We consider 100 independent snapshots of vehicle distribution (with 60\% of cars and 40\% of trucks) in the same environments (i.e., roads and buildings in Fig. \ref{raytracing}), and the CV is randomly selected from a car on the left lane located within $\pm 100$m from the infrastructure.

We assume the infrastructure and the CV both have $N\times N$ uniform planar array antennas, and training beams are generated by the Kronecker-products of $N\times 1$ discrete Fourier transform (DFT) vectors of each horizontal and vertical domain. The 3D beam patterns of $8\times 8$ UPA antennas are shown in Fig. \ref{beampattern}.  Although there are a total number of $N^2 \times N^2$ possible transmit and receive beam pairs to be tested (because there are $N$ DFT vectors in each domain), we only consider the beams that cover the region of interest for exhaustive search. The total numbers of possible training attempts are therefore $23\times 24=552$ for the $8\times 8$ UPA and $53\times 60=3180$ for the $16\times 16$ UPA.

To evaluate the proposed idea, we further assume that the road is divided into grids of 5m distance, and the infrastructure has knowledge of 25 angle-of-arrivals and 25 angle-of-departures to/from each grid when no vehicles are present. This knowledge can be considered as long-term prior channel information because environments are static while instantaneous angle-of-arrivals and angle-of-departures vary due to independent vehicle distribution per snapshot. The infrastructure can predetermine possible transmit and receive training beam pairs for each grid using this prior information. Based on the DSRC signals from the CV, the infrastructure would know the position of the CV and informs the CV about the predetermined training beam pairs and perform beam alignment only using the predetermined training beam pairs. The 5m grid size is selected considering GPS error and the velocity of CV.
\begin{figure*}
\centering
\subfloat[CDF plot of the number of training beams.]{
\includegraphics[width=1\columnwidth]{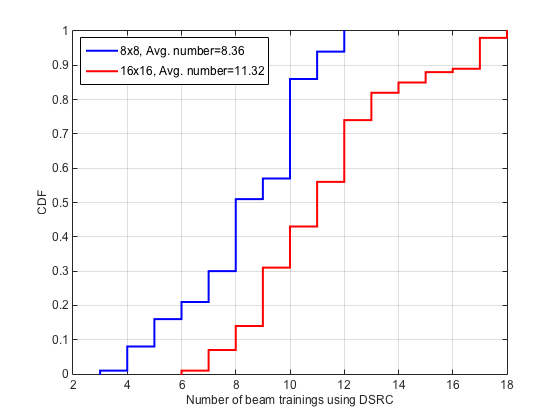}
\label{numbeam}
}
\subfloat[CDF plot of the receiver power loss.]{
\includegraphics[width=1\columnwidth]{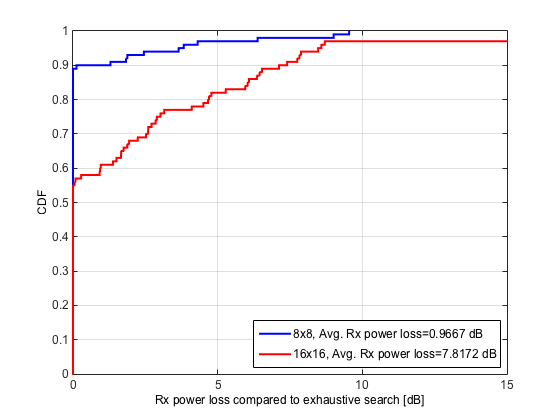}
\label{rxpower}
}
\caption{CDF plots of the number of training beams and the receiver power loss (compared to exhaustive search) of the proposed idea.}
\label{simul_results}
\end{figure*}

Fig. \ref{numbeam} shows the cumulative distribution function (CDF) plots of the number of training beam pairs of the proposed idea. While the numbers of possible training attempts of exhaustive search are 552 and 3180 for the $8\times 8$ and $16\times 16$ UPAs, respectively, the average numbers of training beam pairs of the proposed idea are only around 10 for both cases, which shows the significant reduction of beam alignment overhead.

The proposed idea may have performance degradation compared to exhaustive search due to limited use of training beams without accounting instantaneous traffic conditions. Fig. \ref{rxpower} shows the CDF plots of receive power loss of the proposed idea over exhaustive search. The loss is negligible with the $8\times 8$ UPA. Although the proposed idea suffers from around 8dB loss in average with the $16\times 16$ UPA, more than half of times the proposed idea has negligible performance degradation. It is expected that the loss can be mitigated by developing more advanced beam alignment algorithms exploiting the proposed idea. From Figs. \ref{numbeam}  and \ref{rxpower}, it is clear that the proposed idea can significantly reduce the mmWave beam alignment overhead while having marginal performance degradation.

\section{Conclusions}
In this article, we explained why current technologies for vehicular communications such as DSRC or 4G cellular systems will be insufficient for future connected vehicles that wish to share raw sensor data in large scale. To exchange the large amount of data from automotive sensors, e.g., cameras and LIDARs, we proposed three ways that mmWave can be used in future vehicular networks: 5G cellular, a modified version of IEEE 802.11ad, or a dedicated new standard. We reasoned that a vehicle should have multiple mmWave transceivers to mitigate blockage and have better spatial packing in vehicular environments. We also proposed to use DSRC or automotive sensors as side information to reduce the mmWave beam alignment overhead. We expect exploiting side information obtained from various out-of-band sensors will play a key role not only for mmWave V2X communications but also for all mmWave communication systems in general to achieve sufficient link quality with reduced control overhead.

\bibliographystyle{IEEEtran}
\bibliography{refs_all}

\end{document}